\documentclass[twocolumn,aps]{revtex4}

\begin{document}

\title{Birth of the Universe from the Landscape of String Theory}

\author{Archil Kobakhidze and Laura Mersini-Houghton}

\affiliation{Department of Physics and Astronomy, UNC-Chapel Hill, NC 27517-3255, USA}

\date{\today}

\begin{abstract} 
We show that a unique, most probable and stable solution for the wave function of the universe, with a very small cosmological constant $\Lambda_1 \simeq (\frac{\pi}{l_p N})^2$, can be predicted from the supersymmetric minisuperspace with $N$ vacua, of the landscape of string theory without reffering to the antropic principle. Due to the nearest neighbor tunneling in moduli space lattice, the $N$-fold degeneracy of vacua is lifted and a discrete spectrum of bound state levels over the whole minisuperspace emerges. $SUSY$ is spontaneously broken by these bound states, with discrete nonzero energy levels $\Lambda _s \simeq (\frac{s \pi}{l_p N})^2$, $s = 1,2,..$. 
\end{abstract}

\pacs{xxx}

\maketitle

Recent progress in string theory has revealed a large and rich structure of vacua solutions in moduli space 
\cite{Bousso:2000xa}--\cite{Banks:2003es}, known as the landscape \cite{Susskind:2003kw}. The large number of vacua results from the fact that in a typical compactification of M-theory from eleven dimensions to $
(3+1)$ dimensions there are hundred of ways of 'wrapping' compact dimensions with flux and hundreds of 4-form fields and fluxes. Counting of string theory vacua has been the subject of much recent works on landscape theory\cite{Douglas:2003um}-\cite{Douglas:2004zg}. The (Poincare) supersymmetric (SUSY) vacua are degenerate with zero vacuum energy while the non-SUSY part of the landscape is expected to have  vacua with different but finite values of energy density $\lambda$ in the range $0-M_{p}^4$. On very general ground, one expects to have disconnected sectors of such vacua as well. 

The potential $V(\phi)$ of the moduli field $\phi$, (which collectively describes the contribution from all moduli $\phi_i$), is typically described as having many valleys (the vacua solutions) separated by barriers with height of order $M_p^4$. Though detailed structure of the landscape is not yet fully understood, the large number of vacua solutions will quite likely persist. As a result the following question has been the central theme in the landscape investigation: {\it in which vacuum, from this multitude of choices, does our universe reside}? There seems to be no physical selection criteria to answer this challenge, a fact that has led many people to seek support from anthropic arguments instead \cite{Susskind:2003kw}.

Our approach here is entirely different from the way this challenge has been explored so far. We do not ask 'in which vacuum do we live in?'. Instead, adopting the minisuperspace approach for the second quantization of gravity coupled to moduli fields, we are interested in finding the answer to the following question: {\it which stable solution of the wavefunction of the universe is the most probable solution over the whole superspace of the landscape?} According to the quantum-mechanical description, the universe actually spreads over whole landscape of vacua rather than being relaxed in certain particular vacuum. We model the landscape by considering the superspace of moduli $\phi$ as a finite lattice of $N$ sites with tunneling between sites taken into account in the nearest neghbor approximation. This, as we will show, results in a discrete range of solutions that form 'energy' bands and the most probable one is the minimum energy bound state which is lifted from zero. Thus, the model predicts the universe with small cosmological constant, $\Lambda \simeq (\frac{\pi}{l_p N})^2$ (providing $N$ is large), as the most probable one, without reffering to the antropic arguments.  

\paragraph*{The minisuperspace approach:} 
In the superspace of all vacua solutions for moduli $\phi$ with homogenous $3-$geometries, we confine our study and consider only one sector, namely, the minisuperspace of SUSY vacua, described by  the potential $V(\phi)$ with potential wells that sit at zero, and by the metric of spatially flat and homogenous 3-geometries 
\begin{equation}
ds^2= \left[-{\cal N}dt^2+a^2(t)d\bf{x}^2\right],
\label{1}
\end{equation}  
${\cal N}$ is a lapse function that can be set to ${\cal N}=1$. We also make the assumption of equal {\it a priori} probabilities for each SUSY vacua to occur. This means we consider the potential $V(\phi)$ for the modulus field to have a periodic 'lattice' type distribution of equidistant potential wells and barriers. The energy of SUSY vacua will sit at $\lambda =0$, and for the barrier heights and spacing $b$ between them we can take the typical values of $\cal{O}$$(M_p^4)$ and Planck length $b \simeq l_p$ respectively. Let the number of sites (vacua) in this SUSY minisuperspace 'periodic lattice' be some large but finite number $N$. The SUSY minisuperspace is defined by the configuration space of two variables $\phi$ and $a$.  The wave function of the universe propagating through the SUSY minisuperspace, $\Psi(a,\phi)$, is a functional over configurations $\phi$ and $a$, as we switch on gravity.
All our calculations and results below can easily be extended to closed and open universes.

\paragraph*{Formalism:} We set to solve the Wheeler-De Witt equation in the minisuperspace. The combined  action of $a,\phi$ background is
\begin{equation}
S=S_g+S_{\phi}=\int d^4x\sqrt{-g}\left[\frac{R}{\kappa^2}-\frac{\dot{\phi}^2}{2}-V(\phi)\right]
\label{2}
\end{equation}
We set below the normalization factor $\frac{\kappa^2}{24\pi^2}=\frac{2}{3\pi}M_p^{-2}$  and Planck constant $\hbar$ to be equal to one, unless otherwise noted.
The lagrangian for gravity is
\begin{equation}
L_g=\frac{1}{2}N\left[-\frac{a\dot{a}^2}{N^2}-a^3\Lambda \right]
\label{wheeler}
\end{equation} 
The canonical momenta is defined by $p_a=\frac{\partial L_g}{\partial \dot{a}}=-\frac{a\dot{a}}{{\cal N}}$. The corresponding hamiltonian becomes
\begin{eqnarray}
{\cal H}_g=\left(p_a\dot{a}-L_g\right)/{\cal N}= \nonumber \\
-\frac{1}{2a}\left[p_a^2-a^4\lambda \right]
\end{eqnarray} 

The homogenious moduli field is described by the lagrangian
\begin{equation}
L_{\phi}=\frac{a^3{\cal N}}{2}\left(-\frac{\dot{\phi}^2}{{\cal N}}-V(\phi)\right),
\end{equation}
and its canonical momenta given by $p_{\phi}=\frac{\partial L_{\phi}}{\partial \dot{\phi}}=-a^3\dot{\phi}$.
The potential for the moduli $\phi$ in the {\it SUSY} minisuperspace is some periodic function with zero-energy for all $N$ potential wells and lattice spacing $b$ as explained above, which satisfies
\begin{equation}
V(\phi) = V(\phi + b) 
\label{V}
\end{equation}
with barrier heights $\simeq O(M_{p}^4)$. The hamiltonian for the field is 
\begin{equation}
{\cal H}_{\phi}=\frac{a^3}{2}\left(\dot{\phi}+V(\phi)\right).
\end{equation}
Let us define $ ln(a)=\alpha$, $\frac{\dot{a}}{a}=\dot{\alpha}$. The full Hamiltonian in $(\alpha, \phi)$ becomes
\begin{equation}
{\cal H}=\frac{1}{2e^{3\alpha}}\left[-p_{\alpha}^2+p_{\phi}^2+e^{6\alpha}V(\phi)\right]
\label{4}
\end{equation} 
The system is quantized by promoting conjugate momenta in Eqns. (\ref{3,hp}) to operators $\hat{p}_{\alpha}=-i\frac{\partial}{\partial \alpha}$ and $\hat{p}_{\phi}=-i\frac{\partial}{\partial \phi}$ \footnote{There is the operator ordering ambiguity which can be ignored in the semiclassical approximation we are interested in here. For more detailes and subtleties of the the minisuperspace approach to the second quantization of gravity see, e.g. \cite{Vachaspati:1988as}-\cite{Halliwell:2000mv}.}
\begin{eqnarray}
\hat{{\cal H}}=\frac{1}{2e^{3\alpha}}\left[\frac{\partial^2}{\partial\alpha^2}-
\frac{\partial^2}{\partial\phi^2}+e^{6\alpha}V(\phi)\right] 
\label{wheeler1}
\end{eqnarray} 
The Wheeler-Dewitt equation is the quantum hamiltonian constraint, obtained by varying the action Eqn. (\ref{2}) with respect to ${\cal N}$
\begin{equation}
\hat{{\cal H}}\Psi(a,\phi)=0
\label{wheelerdewitt}
\end{equation}
with the hamiltonian operator ${\cal H}$ given by Eqn. (\ref{wheeler1}) acting on the wave function of the universe $\Psi(\alpha,\phi)$.

The field equations of motion are obtained by varying the action, $S$, Eqn. (\ref{2}) with respect to $\alpha$ and $\phi$, respectively, read
\begin{eqnarray}
\ddot{\phi}+3H\dot{\phi}+\frac{\partial V}{\partial \phi}=0 , \\
\ddot{\alpha}+\frac{3}{2}\left[\dot{\alpha}^2+\dot{\phi}^2-V(\phi)\right]=0 .
\label{4}
\end{eqnarray}   
It is easy to check for consistency that in fact the $\alpha$ equation of motion is nothing more than the Friedman equation for the expansion in the presence of the energy density of the field $\phi$, $ \epsilon = \frac{\dot{\phi}^2}{2}+V(\phi)$

\paragraph*{Boundary Conditions and Solution:} 
The SUSY minisuperspace periodic lattice contains a large but finite number of 'lattice sites' (vacua) $N$ in the potential $V(\phi)$, Eqn. (\ref{V}). We will assume that there is no interaction with the 'hidden sectors' of the superspace and thus the wavefunction does not leak out to other sectors.  The solution to the Wheeler-DeWitt Eqn. (\ref{wheelerdewitt}) would produce an $N-$fold degenerate ground state for $\Psi(\alpha, \phi)$. Let us now allow for tunneling between sites, in the nearest neigbhor approximation, with a tunneling rate $\delta$. Finite large periodic lattices have been extensively studied in condensed matter physics \cite{condensed}. Tunneling between neighbor sites lifts the degeneracy of the vacuum solution as it breaks lattice translation symmetry. In order to establish analogy with condensed matter systems, let us for the moment take $\alpha = constant$ in Eqn. (\ref{wheeler1}) \footnote{Condensed matter systems analogs have often been invoked before, although in different contexts. For a recent analogy to a cascade of phase transitions see e.g. \cite{Kane:2004ct}}. 
For our SUSY 'lattice' the ground state vacuum energy is $\Lambda=0$.
The boundary conditions in this large but finite lattice quantize the wave-number $k$ in terms of a discrete quantum number $s$. There are two possible boundary conditions we can choose for the $N$ lattice sites numbered $0$ to $N-1$ : {\it the fixed end-point} boundary that requires the wave function does not propagate outside the minisuperspace, i.e the function should vanish at these end-points; or the cyclic boundary condition relevant for large $N$ which requires that the $s-th$ site should satisfy $\Psi_{N+s} = \Psi_{s}$. For large N the two are equivalent by symmetry of $k \rightarrow -k, |k|\leq \frac{\pi}{L}$. There are $N-1$ normal modes in this configuration. Boundary conditions thus require that 
\begin{equation}
k_s = \frac{\pi s}{b N}, \ s=1,2,..N.
\label{s}
\end{equation}

Due to the mixing between nearest neighbors from tunneling, the hamiltonian has non-diagonal terms. Diagonalizing the hamiltonian yields the energy eigenvalues of Eqn. (\ref{energy}), thereby splitting the levels and removing the N-fold degenracy of the ground state. The eigenfunctions obtained after the diagonalization of the hamiltonian are the normal modes of the system given by $\Psi_{k(s)}(\phi) \simeq sin(k_{s} \phi)$ or $cos(k_{s} \phi)$. Physically these eigenfunctions are a superposition of left and right moving Bloch plane waves which due to the constructive intereference in their phases satisfy the Bragg reflection condition and form standing waves in the minisuperspace lattice of size $L= b N$. With our boundary condition the standing waves pick the expression
\begin{equation}
\Psi_s \simeq \frac{sin(k_s \phi)}{\sqrt{k_s}}
\label{standing}
\end{equation}
with quantum numbers $s$, (not to be confused with lattice site numering), taking values in the range  $s=1,..N $. The eigenvalues of the hamiltonian form bands of energy with discrete energy levels, $\epsilon_s$.  A rough estimate for the tunneling rate can be given by $\delta \simeq (\frac{\pi}{b})^2$, known as the mass gap of periodic lattices. The energy of each level with wavenumber $k_s$ is
\begin{equation}
\epsilon_s = 2\delta - 2\delta cos(k_{s} b).
\label{energy}
\end{equation}

At this point we re-establish the dependence on $\alpha$ and find solutions for the wavefunction of the universe $\Psi(\alpha, \phi)$ to the full hamiltonian given by Eqn. (\ref{wheeler1}).
Let us take the following ansatz for the wavefunction of the Universe $\Psi$ in Eq. (\ref{5}) 
\begin{equation}
\Psi({\alpha, x})=\Sigma _s  F_{k_s}(\alpha)\psi_{k_s}(x)
\label{6}
\end{equation}
We have rescaled the variable and the parameters as follows: $\phi$ to $x=e^{3\alpha}\phi, \tilde{b} = b e^{3\alpha}, \tilde{k}_s = k_s e^{-3\alpha}, \tilde{\delta}=\delta e^{6\alpha}$ so that Eqn. (\ref{wheelerdewitt}) becomes separable in $\alpha, x$. After rescaling, $\psi_{k_s}(x)$ in Eqn. (\ref{6}) satisfies the $\alpha$ independent equation
\begin{equation}
\left[-\frac{\partial^2}{\partial x^2} + V(x)\right]\psi_{k_s}(x)=\epsilon_{k_s}\psi_k(x)
\label{7}
\end{equation}
The energy eigenvalues $\epsilon_{k_s}\simeq \frac{\hbar ^2 k_{s}^2}{2}$, (as in Eqn. (\ref{energy})), and the solutions for the eigenfunctions $\psi_{k_s}$ are given by Eqns. (\ref{s}, \ref{standing}).

The lowest energy standing wave is the one for $s=1$, $\tilde{k}_1 = \frac{\pi}{\tilde{b} N}$, $\epsilon_1 = (\frac{\pi}{\tilde{b} N})^2$. By plugging Eqn. (\ref{7}) back into Eqn. (\ref{wheelerdewitt}) we obtain that $F_{k_s}(\alpha)$ of Eqn.(\ref{6}) satisfy the following equation
\begin{equation}
\left[\frac{\partial^2}{\partial\alpha^2}+ e^{6\alpha}\epsilon_k\right]F_{k_s}(\alpha)=0.
\label{9}
\end{equation}
The solution to this equation reads
\begin{equation}
F_s(\alpha)\approx \frac{1}{(|\tilde{\epsilon}_s|)^{1/4}}e^{\pm i\sqrt{|\tilde{\epsilon}_s|}}\alpha
\label{10}
\end{equation} 
where $\tilde{\epsilon}_s = e^{6\alpha} \epsilon_s = (\frac{s \pi}{b N})^2$ with $s=1,2, ..N$.

The solution to the equation of motion for $\alpha$, Eqn. (\ref{alpha}) yields $\alpha = \pm|\epsilon_s|^{1/2}t = \pm (H_s t)$. Since the growing mode does soon dominate over the decaying one, we take only the outgoing mode $\alpha = +H_s t$ as our boundary condition at time plus infinity . Therefore, each standing wave mode labelled by the quantum number $s$ in the expression (\ref{6})for the wavefunction $\Psi(\alpha, \phi)$ describes a DeSitter universe with its own constant nonzero cosmological constant $\tilde{\epsilon_s} \simeq (\frac{\pi s}{b N})^2$, time and expansion rate $\alpha = + H_s t$. There are $N-1$ discrete normal modes that form the discrete energy band of bound states, all lifted from zero. Decoherence between levels is resolved since the energy levels are discrete and separated by a finite ammount of energy. The lowest lying energy state, corresponding to $s=1$, has a nonzero energy of $\tilde{\epsilon_1} = (\frac{\pi}{b N})^2$. 

There is ongoing debate in quantum cosmology \cite{Vilenkin:1994rn} on the measure of probability in the wavefunction of the universe, both definitions being plagued with some pathologies. Probability for Eqn. (\ref{wheeler1}), viewed as a Klein Gordon equation is given by: ${\cal P}=i\left(\Psi\partial_a\Psi^{*}-\Psi^{*}\partial_a\Psi \right) $. The same hamiltonian, when treated with the quantum mechanic formalism has a probability given by ${\cal P}=|\Psi|^2$. Due to the oscillatory solution for the modes of $\Psi(\alpha, x)$, in our case both expressions for the probability give, up to an overall normalization constant, ($(\frac{b}{\pi})^2$ for lattices)
\begin{equation}
{\cal P} \approx \frac{1}{|\tilde{\epsilon_{s}}|}
\label{p}
\end{equation}
This shows that ${\cal P}$ is peaked around $\Psi_1$ with energy $\tilde{\epsilon_1} \approx (\frac{\pi}{bN})^2$. Although the SUSY landscape potential has $\lambda=0$, our calculation shows that the most probable solution, Eqn.(~\ref{p}) is peaked around the first bound state in the discrete band of energy levels, i.e the lowest lying energy level $s=1$. As shown the lowest lying level has a non-zero energy  constant energy $\tilde{\epsilon_1} \approx (\frac{\pi}{(bN})^2$. The lifting of the degeneracy of the N-vacua and thus the spontaneous breaking of SUSY  by the bound state $\Psi_1$ due to tunneling gives birth to a Universe with a small cosmological constant $\Lambda =H_1^{2} =\epsilon_1$. $N$ is expected to be large enough. Thus having $\epsilon_1 \approx \Lambda$ in the favoured range of $\Lambda \approx 10^{-120}M_p^4$ can be easily arranged. We have just shown that $\Psi_1$ is a unique, stable and most probable solution with nonzero energy propagating in the SUSY minisuperspace of the landscape. It therefore can be a candidate for the wave function of universe from the landscape.

\paragraph*{Concluding remarks:} We have shown here how a unique, most probable and stable solution for the wave function of the universe can be predicted from the landscape without having to appeal to anthropic arguments. This solution spontaneously breaks SUSY and has a small and non-zero $\Lambda=(\frac{\pi}{l_p N})^2$. Due to tunneling and constructive interference between right and left moving plane waves there are $N-1$ normal mode bound states that occupy the SUSY minisuperspace of the landscape in a discrete band of energy levels that have a non zero mass gap from the degenerate vacuum $\lambda =0$. Each of the bound states have their own energy and can give rise to a DeSitter universe with different expansion rates according to their energy level.  The probability for the bound state levels is inversely proportional to their energy thus the lowest lying energy state becomes the most probable one. It is interesting to note that we get a different estimate for the ground state energy dependence on $N$ , namely we obtained $\Lambda \simeq N^{-2}$ instead of the current estimate appearing in literature i.e. $\Lambda \simeq N^{-1}$. The effect of temperature, nonminimal coupling of moduli to gravity on the lowest energy state and coupling to other sectors, we leave for future work.    

\acknowledgments We are very grateful to Laurie McNeil and Dmitri Khevchenko for many beneficial discussions about condensed matter analogs. We would also like to thank Andreas Albrecht for stimulating  and helpful comments.



\begin{references}

\bibitem{Bousso:2000xa}
R.~Bousso and J.~Polchinski,
``Quantization of four-form fluxes and dynamical neutralization of the
JHEP {\bf 0006}, 006 (2000)
[arXiv:hep-th/0004134].

\bibitem{Kachru:2003aw}
S.~Kachru, R.~Kallosh, A.~Linde and S.~P.~Trivedi,
``De Sitter vacua in string theory,''
Phys.\ Rev.\ D {\bf 68}, 046005 (2003)
[arXiv:hep-th/0301240].

\bibitem{Susskind:2003kw}
L.~Susskind,
``The anthropic landscape of string theory,''
arXiv:hep-th/0302219. 

\bibitem{Freivogel:2004rd}
B.~Freivogel and L.~Susskind,
``A framework for the landscape,''
arXiv:hep-th/0408133.

\bibitem{Banks:2003es}
T.~Banks, M.~Dine and E.~Gorbatov,
``Is there a string theory landscape?,''
JHEP {\bf 0408}, 058 (2004)
[arXiv:hep-th/0309170].

\bibitem{Douglas:2003um}
M.~R.~Douglas,
``The statistics of string / M theory vacua,''
JHEP {\bf 0305}, 046 (2003)
[arXiv:hep-th/0303194].

\bibitem{Denef:2004dm}
F.~Denef, M.~R.~Douglas and B.~Florea,
``Building a better racetrack,''
JHEP {\bf 0406}, 034 (2004)
[arXiv:hep-th/0404257].

\bibitem{Douglas:2004qg}
M.~R.~Douglas,
``Statistical analysis of the supersymmetry breaking scale,'' arXiv:hep-th/0405279.

\bibitem{Douglas:2004zg}
M.~R.~Douglas,
``Basic results in vacuum statistics,''
arXiv:hep-th/0409207.

\bibitem{Vachaspati:1988as}
T.~Vachaspati and A.~Vilenkin,
``On The Uniqueness Of The Tunneling Wave Function Of The Universe,''
Phys.\ Rev.\ D {\bf 37}, 898 (1988).

\bibitem{Vilenkin:1994rn}
A.~Vilenkin,
Phys.\ Rev.\ D {\bf 50}, 2581 (1994)
[arXiv:gr-qc/9403010] and references therein.

\bibitem{Halliwell:2000yc}
J.~J.~Halliwell,
``Trajectories for the wave function of the universe from a simple  detector
model,''
Phys.\ Rev.\ D {\bf 64}, 044008 (2001)
[arXiv:gr-qc/0008046].

\bibitem{Halliwell:2000mv}
J.~J.~Halliwell,
``Approximate Decoherence of Histories and 't Hooft's Deterministic Quantum
Phys.\ Rev.\ D {\bf 63}, 085013 (2001)
[arXiv:quant-ph/0011103].

\bibitem{Kane:2004ct}
G.~L.~Kane, M.~J.~Perry and A.~N.~Zytkow,
arXiv:hep-ph/0408169.

\bibitem{condensed} 
C. Kittel, {\it Introduction to Solid State Physics}, John Wiley \& Sons Inc., 1968.
\end{references}
\end{document}